\documentclass[twocolumn,showpacs,preprintnumbers,amsmath,amssymb,prl]{revtex4}

\usepackage{graphicx}
\usepackage{dcolumn}
\usepackage{bm}
\usepackage{subfigure}

\begin{document}

\title{Shapes of Semiflexible Polymer Rings}

 \author{Karen Alim}
\author{Erwin Frey}
\affiliation{Arnold Sommerfeld Center for Theoretical Physics and Center for NanoScience, Department of Physics,\\ Ludwig-Maximilians-Universit\"at M\"unchen, Theresienstrasse 37, D-80333 M\"unchen, Germany
}%

\date{7 November 2007}

\begin{abstract}
The shape of semiflexible polymer rings is studied over their whole range of flexibility. Investigating the joint distribution of asphericity and nature of asphericity as well as their respective averages we find two distinct shape regimes depending on the flexibility of the polymer. For small perimeter to persistence length the fluctuating rings exhibit only planar, elliptical configurations. At higher flexibilities three-dimensional, crumpled structures arise. Analytic calculations confirm the qualitative behavior of the averaged shape parameters and the elliptical shape in the stiff regime.   
\end{abstract}

\maketitle
It is a well-known fact dating back to 1934 that the shape of a flexible coil is overall prolate \cite{kuhn}. From the isotropy of space, the intuitive expectation would be a spherically symmetric conformation. However, this idea implies rotational averaging and in fact entropy is maximized for a single trajectory of a polymer if the number of segments in each direction is inhomogeneous.

After a series of theoretical investigations, based on both analyses \cite{solc,aronovitz86,diehl} and simulations \cite{bishopclarke, cannon}, only with the onset of single molecule techniques, experiments could prove the asymmetric shape of a flexible polymer \cite{haber00, maier01} and address the relevance of a polymer's shape in biology. The overall shape of a polymer is important for its mobility in heterogeneous media such as cytoplasm and the depletion forces between larger complexes in polymer solution \cite{verma98,triantafillou99}. For the transcription of viral genome or plasmids, the shape of its DNA might enhance or reduce the accessibility for enzymes depending on the spatial distance between DNA-segments \cite{hu06}. DNA as many biopolymers is semiflexible, behaving like a thermally fluctuating elastic rod on length scales of the order of its persistence length. Considering the shape of viral DNA and plasmids this limit is applicable and deserves investigation. In fact, most of the short genomes as well as plasmids are circular, yielding an even stronger constraint for the polymer's shape. To obtain a complete picture for any circular DNA, it is desirable to understand the shape of semiflexible rings as their flexibility is varied. A polymer's shape is well characterized by the asphericity \cite{aronovitz86} as the deviation from spherical symmetry. The degree of prolateness or oblateness is captured by the independent nature of asphericity \cite{cannon}. Measurements of both their mean values give a good indication of how the average outline looks like, but fail to reflect the total ensemble of configurations which can only be accessed via the shape parameter's distribution.  

 We employ Monte Carlo simulations to study the shape of semiflexible polymer rings over a large range of flexibility. To give a complete picture of the polymer's change of shape as its flexibility increases, the joint distribution of  asphericity and nature of asphericity as well as their respective averages will be presented. We find two different shape regimes. In the first, the flexibilities are small resulting in dominantly planar polymer ring configurations. In the second, at large flexibilities, crumpled three-dimensional (3D) structures prevail. In both the stiff and the flexible limit analytic calculations explain the observed behavior.
 
Characterizations of the shape of a polymer's trajectory $\{\mathbf{r}(s)\}$, $s\in[0,L]$, are based on the radius of gyration, primarily a measure for the spatial extent. Generalizing to a radius of gyration tensor $Q$, 
\begin{equation}
Q_{ij}=\frac{1}{L}\int ds\,\mathbf{r}_i(s)\mathbf{r}_j(s)-\frac{1}{L^2}\int ds\,\mathbf{r}_i(s)\int d\tilde{s}\,\mathbf{r}_j(\tilde{s})\,,
\end{equation}
the eigenvalues $\lambda_i$ of the tensor describe the spatial extent along each principal axis.
Measuring the variance of the eigenvalues, the deviation from a fully symmetric object is obtained, denoted asphericity, $\Delta$. Furthermore, prolateness or oblateness of the object is specified by the nature of asphericity, $\Sigma$, measuring the skewness of the eigenvalues. Choosing the normalization such that the quantities are independent of the total length, the asphericity of a polymer is defined by~\cite{aronovitz86}
\begin{equation}
\Delta=\frac{3}{2}\frac{\mathrm{Tr}\,\widehat{Q}^2}{(\mathrm{Tr}\,Q)^2}\;,
\label{eqn_asphericity}
\end{equation}
where $\widehat{Q}_{ij}=Q_{ij}-\delta_{ij}\mathrm{Tr}\,Q/3$. The nature of asphericity is given by~\cite{cannon}
\begin{equation}
\Sigma=\frac{4\mathrm{Det}\,\widehat{Q}}{\left(\frac{2}{3}\mathrm{Tr}\,\widehat{Q}^2\right)^{3/2}}\;.
\label{eqn_nature}
\end{equation}
The asphericity takes values $0\le\Delta\le 1$, where $\Delta=0$ corresponds to a spherically symmetric object. For $\Delta=1$, the polymer is fully extended, forming a rigid rod. The nature of asphericity is bounded between $-1\le\Sigma\le 1$. $\Sigma=-1$ is obtained for a fully oblate object such as a disc, while $\Sigma=1$ is the result for a prolate object as a rigid rod. As the asphericity and the nature of asphericity are independent, a joint distribution yields a thorough classification of stochastic objects such as thermally fluctuating polymers. For reasons of comparison, we will adopt the parameters $\rho=2\sqrt{\Delta}\in[0,2]$ and $\theta=\arccos\Sigma/3\in[0,\pi/3]$ for the joint distribution defined by Cannon et al.~\cite{cannon}.

\begin{figure}[bp]
\centering
\includegraphics[width=0.42\textwidth]{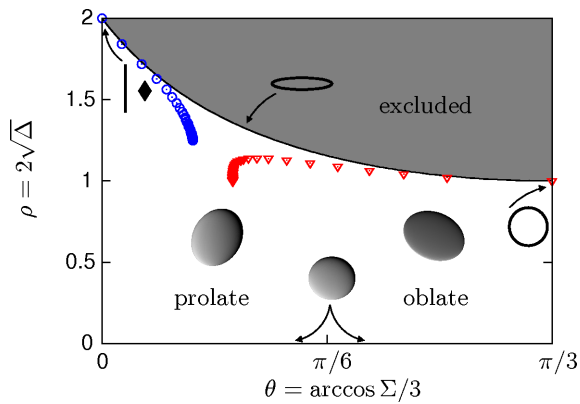}
\caption{(color online). Illustration of the overall shape of polymer configurations depending on the asphericity $\Delta$ and the nature of asphericity $\Sigma$. Along the solid black line the structures are planar; configurations beyond that line are excluded as they do not correspond to real structures. The diamond indicates the peak in the distribution of flexible open polymers. The sequences of circles and triangles denote the mean shape of open and ring polymers at integer flexibilities starting from $L/l_p=0$ at $(0,2)$ and at $(\pi/3,1)$, respectively.}
\label{fig_shapedistroverview}
\end{figure}

Those parameters are directly connected to the eigenvalues of the radius of gyration tensor by: $\lambda_1=\bar{\lambda}(1+\rho\cos(\theta))$, $\lambda_2=\bar{\lambda}(1+\rho\cos(\theta-2\pi/3))$ and $\lambda_3=\bar{\lambda}(1+\rho\cos(\theta+2\pi/3))$, where $\lambda_1\geq\lambda_2\geq\lambda_3$ and $\bar{\lambda}$ denotes the mean eigenvalue. Using these relations, the shape diagram presented in Fig.~\ref{fig_shapedistroverview} is constructed. In the region of both large $\rho$ and large $\theta$ one eigenvalue becomes negative excluding these parameter sets for real structures. Along the solid line separating the excluded conformations from possible ones, at least one eigenvalue is zero. Hence, the solid line represents all planar configurations ranging from the fully oblate geometry of a rigid ring with $\theta=\pi/3$ via elliptical shapes to the fully prolate structure of a rigid rod at $\theta=0$. Below the solid line, 3D conformations are exhibited as all eigenvalues are now greater than zero. The shape is rather oblate for $\theta>\pi/6$ or comparatively prolate for $\theta<\pi/6$ as illustrated by the ellipsoids enclosing a polymer's trajectory. Towards smaller $\rho$, the structure becomes less and less aspherical resulting in a spherically symmetric conformation for $\rho=0$. 

For flexible open polymers the shape distribution is known to be almost exclusively prolate and highly aspherical being peaked around $\theta=\pi/40$ and $\rho=1.55$ \cite{cannon} as indicated by the diamond in Fig.~\ref{fig_shapedistroverview}. However, the conformation of a rigid ring lies just at the opposite end of the shape diagram at $\theta=\pi/3$ and $\rho=1$. As the states of a highly flexible ring polymer can be assumed to be similar to those of flexible open polymers, a strong crossover between a stiff and a flexible regime seems inevitable. Heuristically we may argue that the shape of the tight fluctuating ring in the stiff limit is expected to be dominated by the first modes since higher modes are almost not thermally excited. Both the first in-plane ``breathing'' mode and the first transverse bending mode yield an elliptical conformation as can be illustrated by deforming a strip of paper connected to form a ring. Although the ring rotates in space when fluctuating, the elliptical shape itself remains planar, being oblate for small eccentricities and becoming prolate for large eccentricities of the ellipse. Towards the flexible limit also higher modes are excited resulting in a crossover to the flexible regime where the conformations are three-dimensional and crumpled as expected for a closed random walk.     
 
The METROPOLIS Monte Carlo method was employed to simulate a discretized semiflexible ring of total length $L$ and persistence length $l_p$. The ring is described as a polygon composed of $N$ tethers of fixed length $a=(L/\pi)\sin(\pi/N)$ and direction $\mathbf{t}$. The energy assigned to an individual configuration is given by the elastic energy, $E=Nk_BT(l_p/L)\sum_{i=1}^{N}(1-\mathbf{t}_i\mathbf{t}_{i+1})$, imposing periodic boundary conditions, $\mathbf{t}_{N+1}=\mathbf{t}_1$. New conformations are achieved by pivot moves \cite{klenin}, performing $10^6$ Monte Carlo steps per segment. Measured expectation values of the mean square diameter $\langle D^2\rangle$ were in accordance with analytical expressions \cite{alim} up to the estimated statistical error. The effect of self-avoidance is neglected, as its impact on the shape of even flexible polymers was shown to be only of the order of $1\%$ \cite{aronovitz86}.

\begin{figure}[tp]
\centering
\includegraphics[width=0.38\textwidth]{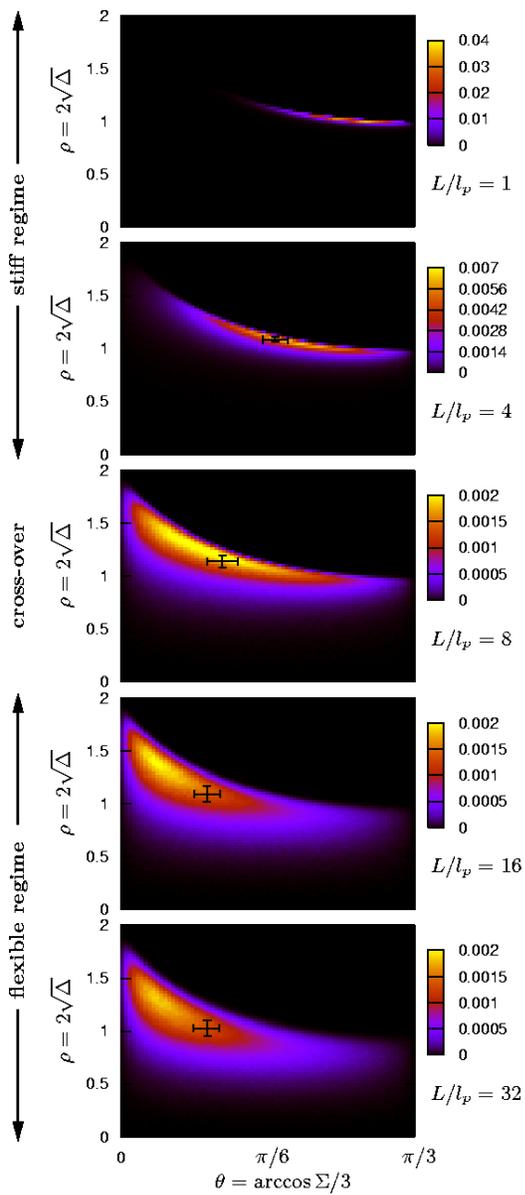}
\caption{(color online). The distribution of asphericity $\Delta$ and nature of asphericity $\Sigma$ at different levels of flexibility. For tight rings, $L/l_p<5$, planar conformations dominate, while the configurations become truly 3D beyond $L/l_p$ of the order of ten. The crosses indicate mean and variance of each single shape parameter deviating from the most probable state. Note the change of the color scaling  as the distribution spreads out.}
\label{fig_shapedistrrhotheta}
\end{figure}

The change of shape as the flexibility increases is best studied when analyzing the shape distribution at different flexibilities as plotted in Fig.~\ref{fig_shapedistrrhotheta}. We distinguish between a stiff regime exemplified by $L/l_p=1,4$ and a flexible regime represented by $L/l_p=16,32$. The geometry of a ring induces an apparent stiffening of the fluctuating polymer to approximately five times its unconstrained flexibility \cite{alim}. Therefore, even $L/l_p=4$ belongs to the stiff limit. By comparison with the shape diagram in Fig.~\ref{fig_shapedistroverview} polymer ring configurations in the stiff regime are identified to be almost exclusively planar ranging from totally oblate to comparatively prolate shapes. In the flexible regime, crumpled structures that fill 3D space dominate the broader configuration space. The distribution changes from rimlike being strongly peaked in the asphericity to lenslike with the major weight on prolate and highly aspherical conformations, although less rodlike than observed for \textit{open} polymers. In agreement with experimental observations \cite{haber00,maier01} the distribution of shapes is very broad yielding also very extended conformation close to $\rho=2$.

In between the two asymptotic shape regimes a crossover is observed represented by $L/l_p=8$ in Fig.~\ref{fig_shapedistrrhotheta}. During this crossover both crumpled, 3D configurations and planar structures are almost equally probable yielding the largest spread of well-occupied conformations in the configuration space.  Beyond the stiff regime in Fig.~\ref{fig_shapedistrrhotheta} the mean of each single shape parameter deviates from the states with largest joint probability, showing its limitation in identifying a polymer's configurations. Identification of the conformations and insight into the width and the form of the density of states are only attainable by the distribution of the shape parameters, highlighting their importance for studying polymer shapes.

\begin{figure}[bp]
\centering
\includegraphics[width=0.45\textwidth]{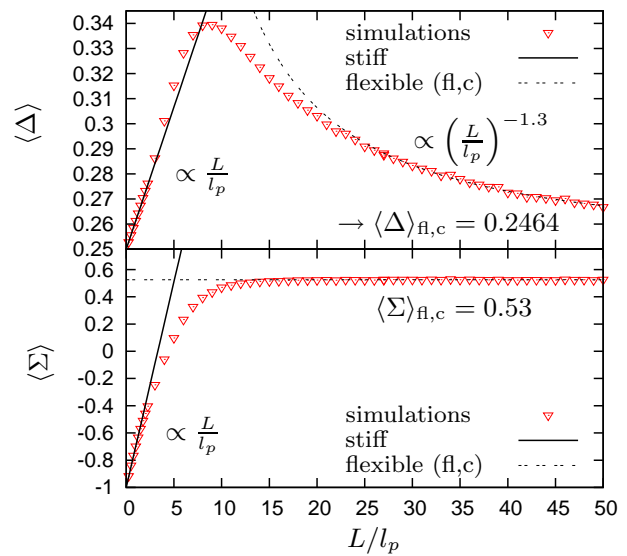}
\caption{(color online). Monte Carlo simulation data for the mean asphericity $\langle\Delta\rangle$ and the mean nature of asphericity $\langle\Sigma\rangle$ versus increasing flexibility $L/l_p$. Both grow linearly with raising flexibility for tight rings (solid line). Then, $\langle\Sigma\rangle$ saturates at a prolate shape, while the asphericity decreases in a power law (dashed line). Error bars are of the size of the symbols.}
\label{fig_3dasp}
\end{figure}
Investigating the mean asphericity and the mean nature of asphericity, qualitative arguments can be quantified and compared to previous results in the limit of infinite flexibility.  The change of both shape parameters on increasing flexibility $L/l_p$ is depicted in Fig.~\ref{fig_3dasp}. For a rigid ring, the asphericity is given by $\Delta=0.25$, being fully oblate: $\Sigma=-1$. Up to $L/l_p\approx 5$ both asphericity and nature of asphericity grow linearly with the flexibility, obeying $\langle\Delta\rangle_{\textrm{stiff}}=0.25+0.01\,L/l_p$, $\langle\Sigma\rangle_{\textrm{stiff}}=-1+0.3\,L/l_p$. This linear dependence classifying the stiff regime is explained by the shape of an ellipse whose axes grow and shrink with the square root of the flexibility, respectively, as will be discussed in the next paragraph. A similar scaling argument has been given by Camacho et al.~\cite{camacho} analyzing planar rings. Beyond this stiff regime a maximum of the mean asphericity is reached. Increasing the flexibility further, higher modes become accessible. These undulations contract particularly the major axis of the ``ellipse'', decreasing the variance of the eigenvalues of the radius of gyration tensor and hence yielding a declining asphericity. The asphericity approaches the exact value for an infinitely flexible polymer ring, a closed Gaussian chain, $\langle\Delta\rangle_{\textrm{fl,c}}=0.2464$, derived by Diehl and Eisenriegler \cite{diehl} in a power law with exponent $\nu=-1.3$. Compared with a flexible open polymer with $\langle\Delta\rangle_{\textrm{fl,o}}=0.396(5)$ \cite{cannon} a polymer ring is much more spherical. Analytic calculations based on a perturbation expansion of a closed Gaussian chain for finite flexibility forecast a positive correction in first order \cite{alim2}, and, hence, explain why the asphericity approaches its Gaussian limit from above.  Also for two-dimensional polymer rings our analytic arguments predict a linear increase of the asphericity in the stiff limit and a monotonic decrease in the flexible limit explaining the nonmonotonic behavior of the shape parameter observed in previous simulations in two dimensions \cite{camacho, norman}. Experiments found a slight increase of the asphericity versus $L/l_p$ in the flexible regime for linear DNA \cite{haber00} not to be justified with our predictions of the shape of semiflexible polymers discarding for the sake of simplicity additional DNA effects, such as twist, nicks or supercoils. Although DNA rings make up a huge field of biological processes where shape matters, resolving their 3D shapes is a challenge due to their small size and rate of change. Ring polymers of larger persistence length $l_p$ such as cytoskeletal filaments as in \cite{claessens} or mesoscopic polymer materials are feasible to measure our results. The nature of asphericity increases monotonically saturating at a value of $\langle\Sigma\rangle_{\textrm{fl,c}}=0.53$ at $L/l_p\approx15$, being less prolate than a flexible open polymers with $\langle\Sigma\rangle_{\textrm{fl,o}}=0.745$ \cite{cannon}. Towards the Gaussian limit the sequences of averaged shape parameters of ring and open polymer as depicted in Fig.~\ref{fig_shapedistroverview} can neither cross nor depart from each other. Therefore, both their $\langle\Sigma\rangle$ approach their limiting value monotonically. Overall, the geometric constraint induces a bias towards more spherical and oblate structures.  

The linear growth of $\langle\Delta\rangle$ and $\langle\Sigma\rangle$ for tight rings is analytically predictable based on the assumption of a planar shape. Due to the first bending modes, the ring of radius $R_c$ becomes an ellipse, where the major and minor axis are the radius $R_c$ increased and decreased, respectively, by $\sqrt{\langle \mathbf{r}_{\perp}^2\rangle}$, the amplitude of the undulations of a weakly bending rod. In the weakly bending limit fluctuations parallel to the average axis of the contour are second order to undulations perpendicular, resulting in the approximate bending energy $E=\frac{1}{2}k_BTl_p\int_0^Lds\:{\mathbf{r}}''_{\perp}(s)^2$, and yielding $\sqrt{\langle \mathbf{r}_{\perp}^2\rangle}=\gamma R _c^{3/2}l_p^{-1/2}$ \cite{camacho, odijkdna,shimada}, where $\gamma$ denotes a numerical constant. Hence, the asphericity and the nature of asphericity are equated in the limit of small flexibilities $L/l_p$, where the first modes truly dominate
\begin{eqnarray}
\Delta_{\mathrm{ellipse}}&=&0.25+2\gamma \,L/l_p+\mathcal{O}((L/l_p)^2)\;,\\
\Sigma_{\mathrm{ellipse}}&=& -1+54\gamma \,L/l_p+\mathcal{O}((L/l_p)^2)\:.
\label{eqn_natureellipse}
\end{eqnarray}
These analytic results forecast the observed behavior.
     
Concluding we have employed the joint distribution of asphericity and nature of asphericity and their respective means as well as analytic arguments to show that the shape of semiflexible polymer rings exhibits two distinct regimes depending on their flexibility. Tight rings are planar ``ellipses'', while flexible rings are 3D, crumpled structures. These two regimes may have implications for a variety of biological processes such as the flow behavior or the accessibility of DNA rings to enzymes. As the shape of stiff, elliptical rings may not be changed by hydrodynamic forces considerably since they will behave as rigid discs, rings in the flexible regime may undergo tumbling motion with alternating collapse and stretching as observed for flexible open polymers \cite{schroeder05}. Similarly, the time it takes an enzyme to find its assigned binding site on a DNA strand should be larger if the DNA conformation is planar, as the enzyme cannot easily travel to DNA-segments separated afar along the backbone by 3D diffusion as in coiled up 3D structures. In this context of opposed behavior in the two shape regimes, polymers in the crossover region where both shapes are equally probable may show striking properties. Depending on the manner a polymer changes  between a planar and a crumpled shape, e.g.~randomly or following a particular trajectory, and its timescale, a broad variety of biological functionality can emerge. The characterization of a semiflexible polymer by its shape can therefore enable a coarse-grained modelling of complex biological processes. 

\acknowledgements
 Financial support of the German Excellence Initiative via the program "Nanosystems Initiative Munich (NIM)" and of the Deutsche Forschungsgemeinschaft through SFB 486 is gratefully acknowledged. 
\acknowledgements

\end{document}